\documentclass[aps,prl,twocolumn,showpacs,preprintnumbers,amsmath,floatfix]{revtex4-1}

\usepackage{graphicx}
\usepackage{epsfig}
\usepackage{dcolumn}
\usepackage{bbm}

\begin{document}
\title{Confinement and the Global $SU(3)$ Color Symmetry}
\author{Ying Chen}
\email{cheny@ihep.ac.cn}
\affiliation{Institute of High Energy Physics, Chinese Academy of Sciences, Beijing 100049, P.R. China}
\affiliation{School of Physics, University of Chinese Academy
of Sciences, Beijing 100049, P.R. China}
\date{\today}
\begin{abstract}
The global $SU(3)$ color symmetry and its physical consequences are discussed. The N\"{o}ther current is actually governed by the conserved matter current of color charges if the color field generated by this charge is properly polarized. The color field strength of a charge can have a uniform part due to the nontrivial QCD vacuum field and the nonzero gluon condensate, which implies that the self-energy of a system with a net color charge is infinite and thereby cannot exist as a free state. This is precisely what the color confinement means. Accordingly, the Cornell type potential with the feature of the Casimir scaling is derived for a color singlet system composed of a static color charge and an anti-charge. The uniform color field also implies that a hadron has a minimal size and a minimal energy. Furthermore, the global $SU(3)$ color symmetry requires that the minimal irreducible color singlet systems can only be $q\bar{q}$, $qqq$, $gg$, $ggg$, $q\bar{q}g$, $qqqg$ and $\bar{q}\bar{q}\bar{q}g$, etc., as such a multi-quark systems can only exist as a molecular configurations if there are no other binding mechanisms.     
\\
\end{abstract}
\maketitle

The color $SU(3)$ symmetry was initially introduced as a global symmetry to solve the spin-statistics problem of baryon wave functions, but caused a new problem that mesons should have color octet partners, which contradicts with experiments. Furthermore, free quarks have never been observed by experiments. This motivates the color confinement conjecture that quarks can exist only in color singlet systems, namely, color singlet hadrons. After the global $SU(3)$ color symmetry is gauged, we obtain the $SU(3)$ gauge field theory\-- the quantum chromodynamics (QCD), which is very successful in explaining the high energy processes and is believed to be the fundamental theory for the strong interaction. Unfortunately, the confinement has not been derived directly from first principles of QCD ever since QCD was established in 1970s. There may be some thing important that is ignored in the conventional treatment of QCD. 

Let us recite the Lagrangian density of QCD of one flavor quark with mass $m$
\begin{equation}
\mathcal{L}_\mathrm{QCD}=-\frac{1}{2}\mathrm{Tr} F_{\mu\nu} F^{\mu\nu}+\bar{\psi}(i D\!\!\!\!/-m)\psi,
\end{equation} 
where $D\!\!\!\!/$ means $\gamma^\mu D_\mu$, $D_\mu=\partial_\mu-igA_\mu$ is the covariant derivative in the presence of the gauge field $A_\mu=A_\mu^a t^a$ with $g$ being the strong coupling, $F_{\mu\nu}=F_{\mu\nu}^a t^a$ is the strength tensor of the gauge field with $F_{\mu\nu}^a=\partial_\mu A_\nu^a-\partial_\nu A_\mu^a+gf^{abc}A^b_\mu A^c_\nu$ ($f^{abc}$ the structure constants of the $SU(3)$ group and are totally antisymmetric with respect to the interchange of the color indexes $a,b,c$. Obviously, $\mathcal{L}_\mathrm{QCD}$ is invariant under the following global transformation 
\begin{equation}
A\to UAU^\dagger, F\to UFU^\dagger, \psi\to U\psi, \bar{\psi}\to \bar{\psi}U^\dagger,
\end{equation}
where $U\in SU(3)$ is an element of $SU(3)$. In other words, the global $SU(3)$ symmetry does exist for QCD. According to the N\"{o}ther's theorem, this continuous global symmetry should result in a conservation law $\partial_\mu j^{a,\mu}=0$, where $j^{a,\mu}$ is the  corresponding N\"{o}ther current 
\begin{equation}\label{current}
j^{a,\mu}=f^{abc}F^{b,\mu\nu}A^{c}_\nu+j_M^{a,\mu} 
\end{equation}
with $j_M^{a,\mu}=\bar{\psi}t^a\gamma^\mu\psi$ being the matter current of quarks. Actually, the current conservation relation is equivalent to the equation of motion of the gauge fields 
\begin{equation}\label{eom1}
([D_\mu, F^{\mu\nu}])^a\equiv \partial_\mu F^{a,\mu\nu}+gf^{abc}F^{b,\nu\mu}A_\mu^c=-gj^{a,\nu}_M.
\end{equation}
One can see this by acting $\partial_\nu$ on the two sides of the above equation. These are well-known results. Conventionally, the N\"{o}ther current and its physical consequence have been seldom discussed seriously since $j^{a,\mu}$ depends explicitly on the gauge field $A_\mu^a$ which is gauge dependent. 

However, this is not the whole story for that the non-Abelian characteristic of $SU(3)$ depicted by the antisymmetric structure constants $f^{abc}$ has more implications. Imagine that a quark current of $a_0$ type, namely $j_M^{a_0,\mu}$, is put into the vacuum, if the color field it generates through the field equation Eq.(\ref{eom1}) has the property
\begin{equation}\label{constraint}
A_\mu^a(x;j^{a_0}_M)=V_\mu(x)\delta^a_{a_0},
\end{equation}
then the first term of the total N\"{o}ther current in Eq.~(\ref{current}) vanishes owing to the asymmetric $f^{abc}$ and one has $j^{a,\mu}=j^{a_0,\mu}_M \delta^a_{a_0}$, which is free of the gauge dependence. Under this condition, it is easy to prove that the corresponding N\"{o}ther charge $Q^a=\int d^3 \mathbf{x}j^{a,0}=\delta^a_{a_0}\int d^3\mathbf{x}j^{a_0,0}_M$ is conserved, namely,
\begin{equation}\label{law}
\frac{d}{dt}Q^a=-\int d^3\mathbf{x}\nabla\cdot \mathbf{j}^{a}=-\delta^a_{a_0}\int d^3{\mathbf{x}}\nabla\cdot \mathbf{j}_M^{a_0}=0,
\end{equation}
if the boundary condition $\psi(\mathbf{x}= \infty,t)=0$ is assumed. Alternatively, we can treat the condition Eq.~(\ref{constraint}) as an implicit constraint in order for the N\"{o}ther
current to be physically meaningful. The direct consequence of this constraint is that the color current of matter fields $j_M^{a,\mu}$ is conserved. 
 
With the constraint Eq.~(\ref{constraint}), the equation of motion Eq.~(\ref{eom1}) is simplified as 
 \begin{equation}\label{eom2}
 \partial_\mu f^{\mu\nu}=-gj_M^{a_0,\nu},
 \end{equation}
 where $f_{\mu\nu}=\partial_\mu V_\nu-\partial_\nu V_\mu$. The Bianchi identity also gives 
 $\partial_\mu \tilde{f}_{\mu\nu}=0$ with $\tilde{f}_{\mu\nu}=\frac{1}{2}\epsilon_{\mu\nu\rho\delta}f^{\rho\delta}$. These equations are very similar to the Maxwell's equations of QED. The solution of Eq.~(\ref{eom2}) can be written as 
\begin{equation}\label{solution}
f_{\mu\nu}=f_{1,\mu\nu}+f_{0,\mu\nu},
\end{equation}
where $f_{1,\mu\nu}$ is a special solution of Eq.~(\ref{eom1}) and $f_{0,\mu\nu}$ is the general solution of the 
homogeneous equation $\partial_\mu f^{\mu\nu}=0$. If we are constricted to the static case, then $f_{0,\mu\nu}$ can has the following form 
  \begin{equation}\label{strength0}
 f_{0,\mu\nu}\propto g\sigma (n_{1,\mu} n_{2,\nu}-n_{2,\mu} n_{1,\nu}),
 \end{equation}
where $\sigma$ is a constant of dimension two, $n_1^\mu$ and $n_2^\nu$ are two constant Lorentz unit vectors, and the gauge coupling appears here is just a convention. Because we do not know how the color charge is quantified at present, we use the proportional sign in the above and the following relations. Consequently, the scalar product of $f_{0,\mu\nu}$ can be 
expressed in terms of $n_{1,2}$ as
\begin{equation}
f_{0,\mu\nu}f_0^{\mu\nu}\propto 2g^2\sigma^2\left[n_1^2 n_2^2-(n_1\cdot n_2)^2 \right].
\end{equation}
For the case of $n_1^\mu$ and $n_2^\mu$ being light-like, say, $n_1^2 =n_2^2 =0$ and satisfy $n_1\cdot n_2=1$, we have 
 $f_{0,\mu\nu}f_0^{\mu\nu}\propto -2g^2\sigma^2$.

In order to determine the solution Eq.~(\ref{strength0}), we would like to consider the proper boundary conditions. Obviously, the discussion above also applies to QED. In the static case, the electromagnetic field strength $F^{\mu\nu}_\mathrm{QED}(\mathbf{x})$ behaves as $F_{\mu\nu}(\mathbf{x}) \sim 1/|\mathbf{x}|^2$ when $|\mathbf{x}|\to \infty$, such that it is natural to set $\sigma=0$ for QED. However, for a non-Abelian gauge field theory such as QCD, the vacuum is highly nontrivial and there is no {\it ab initio} requirement that $F_{\mu\nu}(\mathbf{x})$ vanish at the infinite point, therefore $\sigma$ can take a non-zero value. Actually, the highly nontrivial QCD vacuum is reflected by the topology, the trace anomaly, and the non-zero condensates of quarks and gluons, for example, the gluon condensate $\langle \alpha_s F_{\mu\nu}^a F^{a,\mu\nu}\rangle \ne 0$, which permit the possibility of a non-zero $\sigma$. 

On the other hand, one can find $F_{\mu\nu}\propto f_{0,\mu\nu}\sum\limits_a t^a$ is a non-trivial solution of the vacuum field equation $[D_\mu,F_{\mu\nu}]=0$, and is uniform in both the space-time and the color space. The vacuum expectation value of the scalar product of this field is $\langle \mathrm{Tr} F^2\rangle \propto -8 g^2 \sigma^2$ which means that $\sigma$ has a direct connection with the gluon condensate. Thus the field $f_{0,\mu\nu}$ generated by the color current $j_M^{a_0,\mu}$ can be thought of that the vacuum field $F_{\mu\nu}$ is uniformly polarized to the $a_0$ type through 
$f_{0,\mu\nu}\propto 2 \mathrm{Tr} F_{\mu\nu}t^{a_0}$.

\begin{figure}[t!]
	\includegraphics[height=6.0cm]{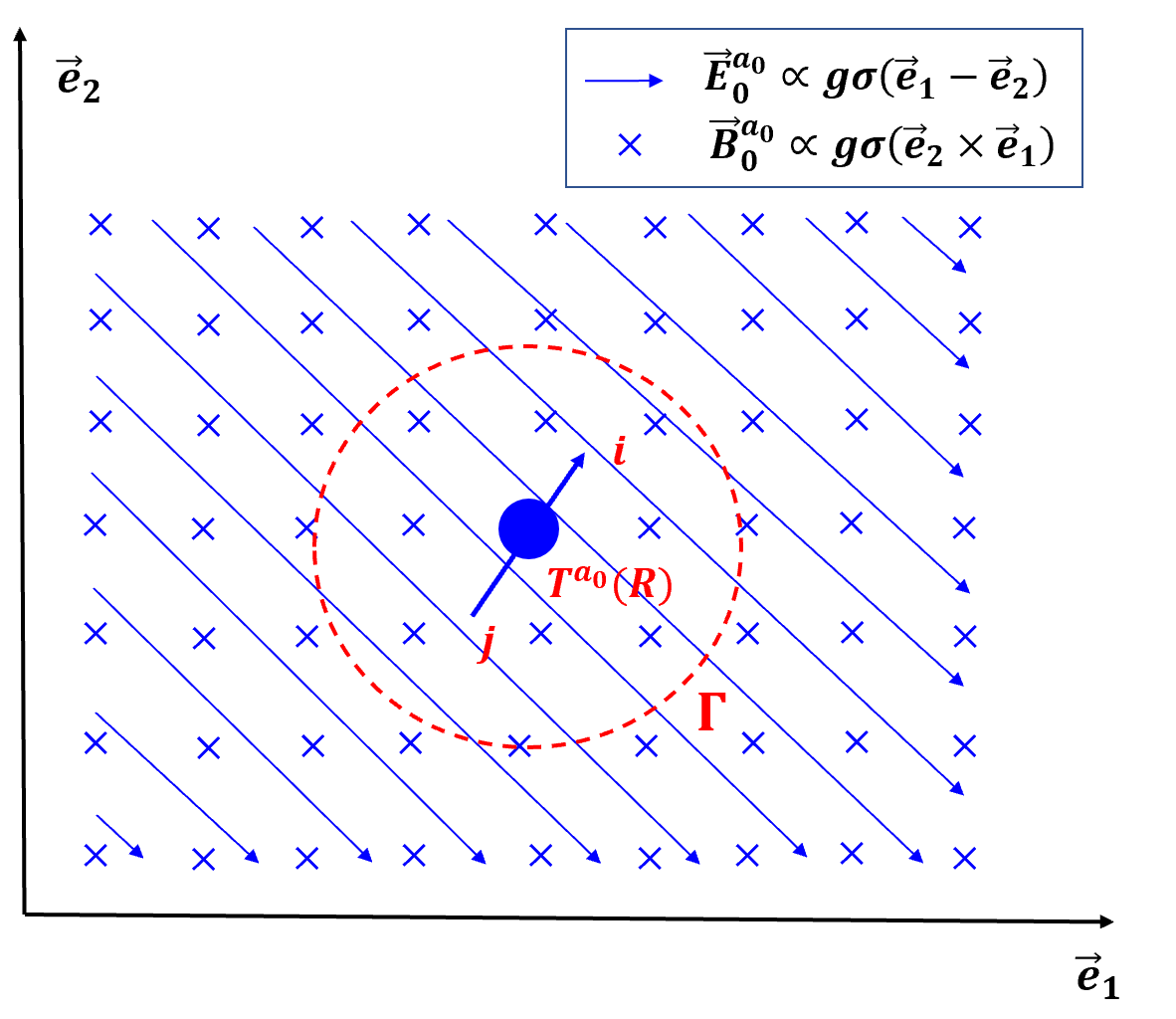}\quad
	\caption{\label{fig:field} Schematic plot the distribution of the uniform color field $\mathbf{E}^{a_0}_0$ and $\mathbf{B}^{a_0}_0$ generated by the charge $T^{a_0}(R)$. The arrowed straight lines illustrate the uniform $\mathbf{E}^{a_0}_0$ oriented in the $\mathbf{e}_1-\mathbf{e}_2$ direction, while '$\times$''s shows the uniform $\mathbf{B}^{a_0}_0$ oriented in the $\mathbf{e}_2\times\mathbf{e}_1$ direction. The indecies $i,j$ stand for the components of the color charge in the $R$ representation. The dashed circle $\Gamma$ illustrates an arbitrary closed surface surrounding the charge. It is easy to see that the total field flux outward from $\Gamma$ vanishes.}
\end{figure}

Now let us consider the free energy of a color charge. In the rest frame of a color charge of type $a_0$, $n_1^\mu$ and $n_2^\mu$ can be chosen to be $n_1=(1,\mathbf{e}_1)$ and $n_2=(1,\mathbf{e}_2)$, where $\mathbf{e}_{1,2}$ are arbitrary spatial unit vectors with $\mathbf{e}_1\perp \mathbf{e}_2$. According to Eq.(\ref{strength0}), we have the corresponding uniform chromo-electric and chromo-magnetic field strengths
\begin{equation}\label{strength-ex}
\mathbf{E}^{a_0}_0 \propto g\sigma(\mathbf{e}_1-\mathbf{e}_2),~~~ \mathbf{B}^{a_0}_0\propto g\sigma(\mathbf{e}_2\times \mathbf{e}_1),
\end{equation}
which obey the normal parity transformation property and $\mathbf{E}^{a_0}_0\cdot\mathbf{B}_0^{a_0}=0$ (no $CP$ violation). Their distribution is illustrated in Fig.~\ref{fig:field}. These expressions are in exact agreement with the conservation of color charge in that for an arbitrary closed spatial surface $\Gamma$
surrounding the color charge, the total flux of the polarized vaccum field outward from it is zero. We use $\mathbf{E}^{a_0}_1$ and $\mathbf{B}^{a_0}_1$ to denote the field strength given by the solution $f_{1,\mu\nu}$ in Eq.~(\ref{solution}), thus the energy of the color field generated by the color charge 
is 
\begin{equation}
E_f\propto\int d^3\mathbf{x} \frac{1}{2}\left[ (\mathbf{E}^{a_0}_0+\mathbf{E}^{a_0}_1)^2+ (\mathbf{B}^{a_0}_0+\mathbf{B}^{a_0}_1)^2\right],
\end{equation}
which is by definition the free energy of the color charge. Note that  $\mathbf{E}^{a_0}_1$ and $\mathbf{B}^{a_0}_1$ behave as $O(1/|\mathbf{x}|^2)$ when $|\mathbf{x}|\to \infty$, the leading term of $E_f$ is $\frac{3}{2} g^2\sigma^2 V$ where $V\to \infty$ is the total spatial volume (It is understood that $\mathbf{E}_1^{a_0}$ and $\mathbf{B}_1^{a_0}$ are properly regularized in the ultraviolet region). In other words, the free energy of an isolated color charge is infinite large.  Lattice QCD confirms this by the observation that the vacuum expectation value of the Polyakov loop $L$ is zero in the confinement phase, say, $\langle L\rangle \sim e^{-E_fT}=0$, where $T$ is the temporal extension of the lattice. Therefore isolated color charges, such as quarks and color octet mesons, do not exist. This is exactly the color confinement argument. The above discussion can be extended simply to color field generated by the current $j^{a,\mu}t^a$, whose strength is $F_{\mu\nu}\propto (f_{1,\mu\nu}+f_{0,\mu\nu})\sum\limits_a t^a$.

We give some comments on the discussion above: First, the color charge providing the color current $j^{a,\mu}_M$ can be point-like or be a spatially extended object composed of several 
subsystems of the irreducible representations (irreps) $R_1, R_2,\ldots$. According to the representation theory of $SU(3)$ group, 
\begin{equation}
R_1\otimes R_2\otimes \cdots \to R_1'\oplus R_2' \oplus R_3'\oplus \cdots,
\end{equation} 
the representation of the color charge can be anyone of the irreps $R'_i$. The global color symmetry and the conservation law of Eq.~(\ref{law}) requires the whole system keeps its color state of $R'_i$ regardless of however complicated the microscopic internal structure is. Except for the color singlet, the vacuum will be polarized with respect to the current of $R'_i$, such that the free energy of the color charge is infinite. Secondly, a color singlet system does not provide a color current and therefore does not interaction with the vacuum color field. It can move freely in the unpolarized color vacuum. 
 \begin{figure}[t!]
	\includegraphics[height=6.0cm]{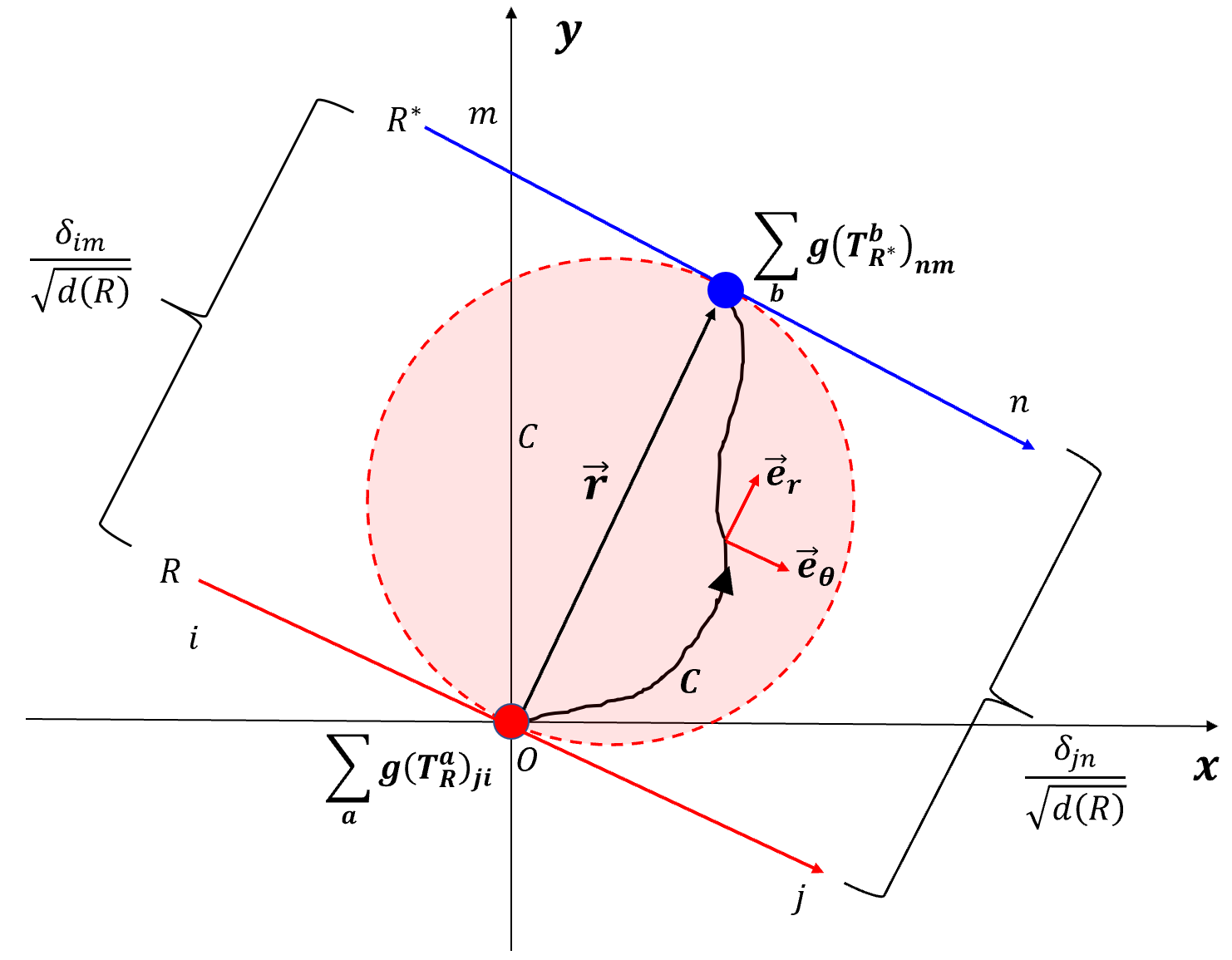}\quad
	\caption{\label{fig:potential} The interaction of color charges of representations $R$ and $R^*$ in a color singlet. Imagine that $R$ and $R^*$ are produced at the origin $O$, then $R^*$ moves to $\mathbf{r}$. The initial and final color states of $R(R^*)$ are labeled as $i,j (m,n)$, respectively, thus the color wave functions of the initial and the final state of singlet $RR^*$ are $\frac{\delta_{im}}{\sqrt{d(R)} }$ and $\frac{\delta_{jn}}{\sqrt{d(R)} }$. If we choose $\mathbf{e}_1=\mathbf{e}_r$ and $\mathbf{e}_2=\mathbf{e}_\theta$, then the uniform chromo-electric field does work $\Delta E=-C_2(R)g^2\sigma r$ regardless of the path $C$ along which $R^*$ moves from $O$ to $\mathbf{r}$. }
\end{figure}

Since color charges exist only in color singlets, we now consider the simplest case that a color singlet system which is composed of a color charge in the $R$ representation and an anti-color charge in the conjugate representation $R^*$ of $R$ with the dimension of the representation is $d_R$. We can treat this system as the $R^*$ charge scatters against $R$ charge through the color field generated by the $R$ charge, as shown in Fig.~\ref{fig:potential}. If the initial and the final state of the $R$ charge has the color indexes $i,j$, respectively, with $i,j=1,2,\ldots, d_R$, then the color charge of $R$ is $g\sum\limits_a (T_R^{a})_{ji}$, where $T_R^{a}$ is the representation matrix of the generators of $SU(3)$ in the $R$ irreps. Similarly, the color charge of $R^*$ is $g\sum\limits_b (T_{R^*}^{b})_{nm}$ with $m,n=1,2,\ldots, d_R$ being the indexes of the initial and final states of $R^*$ charge. With regard to the relation $T_{R^*}^a=-(T_R^a)^*$ and $T^{a}_R=T^{a,\dagger}_R$, the color singlet 
requires that the their interaction should be proportional to 
\begin{eqnarray}
&&\sum\limits_{a,b} (T_R^a)_{ji}(T_{R^*}^b)_{nm}\frac{\delta_{im}}{\sqrt{d_R}}\frac{\delta_{jn}}{\sqrt{d_R}}\nonumber\\ &=&-\frac{1}{d_R}\sum\limits_{a}(T_R^a)_{ji}(T_{R}^{a})_{ij}= -C_2(R)
\end{eqnarray}
where $C_2(R)$ is the eigenvalue of the second order Casimir operator of the $SU(3)$ group in the $R$ irreps and the minus sign implies that the interaction between the $R$ and $R^*$ charge is attractive. In QED, the equation of motion of an electric charge $q$ in an external field $F^{\mu\nu}_\mathrm{QED}$ is $\frac{dp^\mu}{d\tau}=qF^{\mu\nu}_\mathrm{QED}u_\nu$ where $p^\mu=(E,\mathbf{p}_\mathrm{kin})$ is the energy-momentum vector of the charge, where $\mathbf{p}_\mathrm{kin}$ is the kinetic momentum, $u^\mu=\frac{dx^\mu}{d\tau}$ is its velocity vector. Similarly, the motion of equation of the $R^*$ charge in the field of the $R$ charge can be written as 
 \begin{equation}
 {dp^\mu}\propto -g C_2(R) (f_{1}^{\mu\nu}+f_0^{\mu\nu})dx_\nu.
 \end{equation}
In the rest frame of the $R$ charge with the $R$ charge being at the origin,  we can take $u^\mu=(1,\mathbf{v})$. If we ignore $f_1^{\mu\nu}$ temporarily, according to Eq.~(\ref{strength0}) and (\ref{strength-ex}), we have 
\begin{equation}
dE=-C_2(R)g^2\sigma(\mathbf{e}_1-\mathbf{e}_2)\cdot d\mathbf{r}. 
\end{equation}
Based on the discussion above, $\mathbf{e}_1$ and $\mathbf{e}_2$ are originally arbitrary spatial unit vectors. As shown in Fig~\ref{fig:potential}, if we imagine that the $R^*$ charge moves from the origin to the point $\mathbf{r}$, then $\mathbf{e}_r$ is a special direction. Thus we can choose $\mathbf{e}_1=\mathbf{e}_r$ and $\mathbf{e}_2=\mathbf{e}_\theta$ with $\mathbf{e}_\theta \perp \mathbf{e}_r$. If $R^*$ moves from the origin $O$ to $\mathbf{r}$ through an arbitrary path $C$, then the energy change is 
\begin{equation}
\Delta E=-C_2(R)g^2\sigma \int\limits_{\mathbf{0},C}^{\mathbf{r}}d\mathbf{r}'\cdot(\mathbf{e}_r-\mathbf{e}_\theta)=-C_2(R)g^2\sigma r
\end{equation}
where $r=|\mathbf{r}|$ and the proportional relation is replaced by the equal relation with $\sigma>0$ being a parameter to be determined. This means that the chromo-electric field does negative work. The conservation of energy requires the (kinetic) energy loss is reserved in other types of energy. Since $\Delta E$ is independent of the path $C$, we can interpret this energy type is the potential energy $V(r)$. Now taking into account the Coulomb potential energy generated by the point-like color charge $R$, the total potential is 
\begin{equation}\label{cornell}
V(r)=V_0-C_2(R) \frac{\alpha_s}{r}+C_2(R)g^2\sigma r,
\end{equation}
where $\alpha_s =\frac{g^2}{4\pi}$. This is exactly the Cornell potential. It is very interesting to see that the potential has a property of 'Casimir scaling' that the coefficients of the Coulomb part and the linear part are proportional to $C_2(R)$. The Cornell potential and the Casimir scaling have been observed for a long time by the lattice QCD calculations~\cite{bali}. Here we have given the first decent derivation of them. 

The effect of the chromo-magnetic field on the color charge $R^*$ is very similar to a uniform
magnetic field on an electric charge. According to Eq.~(\ref{strength-ex}), the chromo-magnetic field is oriented perpendicularly to the $(\mathbf{e}_r,\mathbf{e}_\theta)$ plane, such that it does not do work to the charge $R^*$. With the present of the chromo-magnetic field, the kinetic momentum $\mathbf{p}_{\mathrm{kin}}$ is related to the canonical momentum $\mathbf{p}$ and the vector potential as 
\begin{equation}
\mathbf{p}_{\mathrm{kin}}=\mathbf{p}+g C_2(R)\mathbf{V}.
\end{equation}  
where $\mathbf{V}$ is the spatial component of $V_\mu$. According to Eq.~(\ref{strength0}) and $f_{\mu\nu}=\partial_\mu V_\nu-\partial_\nu V_\mu$, one has
\begin{equation}
V_{\mu}=\frac{1}{2}g\sigma\left[n_1\cdot (x-x_0) n_{2,\mu}-n_2\cdot (x-x_0) n_{1,\mu}\right].
\end{equation}
If the charge $R^*$ is at $\mathbf{r}$, where $x^\mu =(0,\mathbf{r})$ and $x_0^\mu=(0,\mathbf{0})$, one has
\begin{equation}
\mathbf{V}=\frac{1}{2}g\sigma y \mathbf{e}_x-\frac{1}{2}g\sigma x \mathbf{e}_y
\end{equation} 
where $\mathbf{e}_x$ and $\mathbf{e}_y$ are the orientation vectors of the $x$-axis and $y$-axis in the $(\mathbf{e}_r,\mathbf{e}_\theta)$ plane. Thus the $p_{\mathrm{kin},x,y}$ and $p_{x,y}$ are related as 
\begin{eqnarray}
p_{\mathrm{kin},x}=p_x+\frac{1}{2}g^2C_2(R)\sigma y\nonumber\\
p_{\mathrm{kin},y}=p_y-\frac{1}{2}g^2C_2(R)\sigma x
\end{eqnarray}
Considering the quantum effects, the momentum and the coordinate are operators and satisfy the 
canonical commutation relation $[x_i,p_i]=i$. When the kinetic momentum is much smaller than 
the potential in magnitudes, say, $|\mathbf{p}_{\mathrm{kin}}|\ll |gC_2(R)\mathbf{V}|$, we have 
\begin{equation}
[x,y]=-i\frac{2}{C_2(R)g^2\sigma},
\end{equation}
which means that the coordinates in the $(\mathbf{e}_r,\mathbf{e}_\theta)$ plane are non-commutative and have the uncertainty relation
\begin{equation}
\Delta x \Delta y\sim \frac{2}{C_2(R)g^2\sigma}.
\end{equation}
The uncertainty relation implies that a hadron has a minimal projected area $\frac{2}{C_2(R)g^2\sigma}$ in the $(\mathbf{e}_r,\mathbf{e}_\theta)$ plane and thereby has a minimal size. According to lattice QCD and phenomenological results of the string tension of the $Q\bar{Q}$ system
\begin{equation}
\sigma_{\rm Q\bar{Q}}=\frac{4}{3}g^2 \sigma \sim 0.2 (\mathrm{GeV})^2,
\end{equation} 
the minimal size of a hadron can be estimated as  
\begin{equation}
r_{\rm h}\sim \sqrt{(\Delta x)^2 +(\Delta y)^2}\ge\sqrt{2\Delta x \Delta y}\sim\frac{2}{\sqrt{\sigma_{Q\bar{Q}}}}\sim 0.89~\mathrm{fm}.
\end{equation} 
This minimal size subsequently gives an estimate of the minimal enery of a hadron. Considering a color singlet $q\bar{q}$ system of a spatial volume $V$ and referring to Eq.~(\ref{cornell}), the mean charge charge of 
the (anti) quark can be taken as $\sqrt{C_2(F)}$, where $C_2(F)=4/3$. If the Coulomb part is 
ignored again, the color field strength around the (anti) quark is 
\begin{equation}
\mathbf{E}_0=g\frac{2}{\sqrt{3}}\sigma(\mathbf{e}_1-\mathbf{e}_2),~~~ \mathbf{B}_0=g\frac{2}{\sqrt{3}}\sigma(\mathbf{e}_2\times \mathbf{e}_1),
\end{equation} 
such that the minimal energy in the volume is 
\begin{equation}
E_\mathrm{min}=\int\limits_V d^3 \mathbf{x} \frac{1}{2}(\mathbf{E}_0^2+\mathbf{B}_0^2)\sim \frac{3\sqrt{\sigma_{Q\bar{Q}}}}{8\alpha_s} 
\end{equation}
which gives $E_\mathrm{min}\sim 550$ MeV if $\alpha_s \sim 0.3$. This is a very rough but somewhat reasonable estimate. 



\begin{figure}[t!]
	\includegraphics[height=6.7cm]{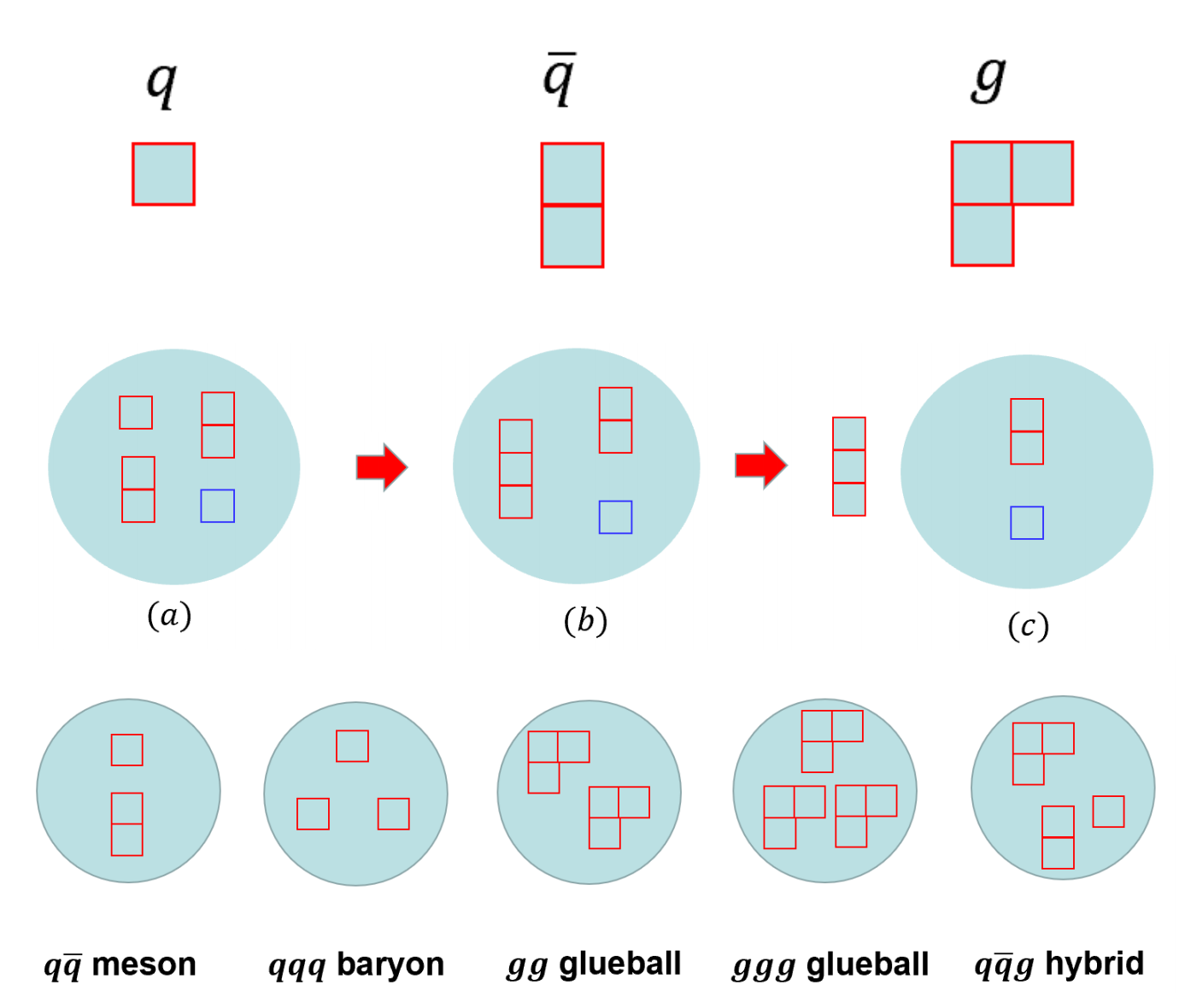}\quad
	\caption{\label{fig:structure} The first row: the Young diagrams for a quark $q(\mathbf{3})$, an antiquark $\bar{q}(\mathbf{3}^*)$ and a gluon $g(\mathbf{8})$. The second row: The diagram ($a$) represents a $q\bar{q}q\bar{q}$ color singlet system. The diagram ($b$) implies that for an objective quark (the blue block), the rest parts act as a $\mathbf{3}^*$ charge composed of an antiquark and a color singlet $q\bar{s}$. The diagram $(c)$ shows that the color singlet $q\bar{q}$ escapes from the system, such that the objective quark feels only the field generated by the antiquark and results in a $q\bar{q}$ meson. The third row shows some of the minimal irreducible color singlets composed of (anti)quarks and gluons in the meaning of the global $SU(3)$ color symmetry.}
\end{figure}

The global $SU(3)$ color symmetry has further implications. Since an isolated color charge does not exist, it is reasonable to assume that the whole universe is in a color singlet. Therefore, when an objective color charge of irrep $R$ is considered, one must keep in mind that it is not isolated but coexists with a system, namely, the residual world, of the conjugate irrep $R^*$. This is somewhat a kind of color interference or color entanglement. The global symmetry requires that however complicated the residual world is, its effects on the objective charge $R$ are equivalent to that of a charge $R^*$. The logic is illustrated in Figure~\ref{fig:structure}. The fundamental degrees of freedom of QCD are quarks, antiquarks and gluons, which are the $\mathbf{3}$, $\mathbf{3}^*$ and $\mathbf{8}$ irreps of the color $SU(3)$ group, as expressed by the Young diagrams in the first row of Fig.~\ref{fig:structure}. For a color singlet $q\bar{q}q\bar{q}$ system (diagram $(a)$ in the second row), the rest part of the system (red blocks) except for an objective quark (blue block) acts as a $\mathbf{3}^*$ charge composed of an antiquark and a singlet $q\bar{q}$ block, as described by diagram ($b$). The diagram ($c$) says that the color singlet $q\bar{q}$ can escape from the system freely and the residual $q$ and $\bar{q}$ compose a color singlet $q\bar{q}$ meson. This discussion can be extended to any a color singlet system made up of many colored objects, which can be reduced iteratively following the above logic. In this sense, the possible irreducible color singlet
configurations are $q\bar{q}$ (meson), $qqq$ (baryon), $\bar{q}\bar{q}\bar{q}$ (anti-baryon), $q\bar{q}g$ (hybrid meson), $qqqg$ (hybrid baryon), $gg$ and $ggg$ (glueballs), etc., as shown in the third row of Fig.~\ref{fig:structure}. If we treat the components of the irreducible color singlets list above as constituent (anti)quarks
and gluons, then the number of them might be a good quantum number in the picture of the global color symmetry. As a result from above, if there is no other mechanisms, compact multiquark configurations are disfavored except for hadronic molecules bound by the residual strong interaction between color singlet objects. 

To summarize, the global $SU(3)$ color symmetry has physical significance.  If the color field generated by a color current of type $a_0$ satisfies $A_\mu^a=V_\mu\delta_{a_0}^a$, then the N\"{o}ther current is gauge independent and is actually governed by the matter current of this charge which is conserved in the conventional meaning. There does exist a solution to the equation of motion of the color field under this constraint, whose strength has a uniform part owing to the nontrivial QCD vacuum and the nonzero gluon condensate. Resultantly, a system with a net color charge has a infinite large free energy and thereby cannot exist freely, as required by the color confinement argument. Following this logic, the potential between  a static color charge and an anti color charge is derived to be exactly the Cornell type, and satisfies the Casimir scaling observed by lattice QCD. For a color singlet system, such as a hadron, the uniform chromo-magnetic field also implies that the system has a minimal size and a minimal energy. Furthermore, since quarks and gluons are fundamental degrees of freedom of QCD, the global $SU(3)$ color symmetry requires that the possible irreducible color singlet systems can only be $q\bar{q}$, $qqq$, $gg$, $ggg$, $q\bar{q}g$ and $qqqg$, etc., such that a bound multi-quark system beyond the above configurations can only exist as a hadron molecule, if there are no other binding mechanisms.     

\par
 This work is supported of the National Science Foundation of China (NNSFC) under Grants No.11935017, No. 11575196, No. 11621131001 (CRC 110 by DFG and NSFC) and the Strategic Priority Research Program of Chinese Academy of Sciences (No. XDB34030302). The author also acknowledges the support of the CAS Center for Excellence in Particle Physics (CCEPP).

\end{document}